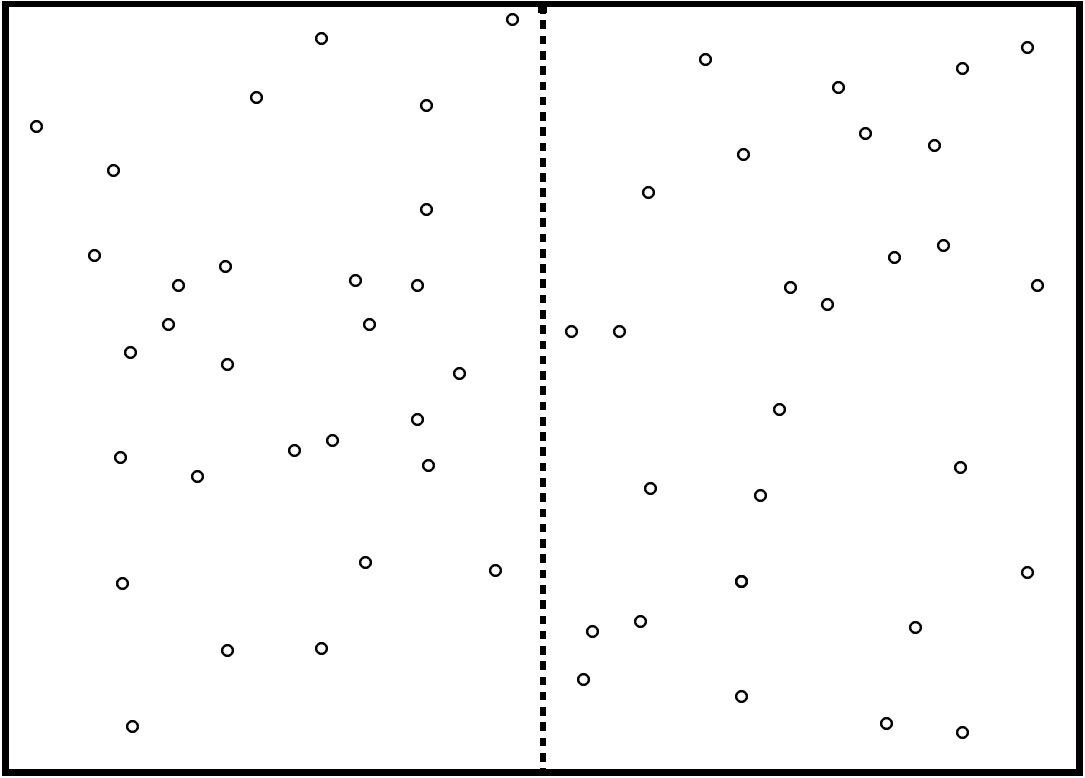

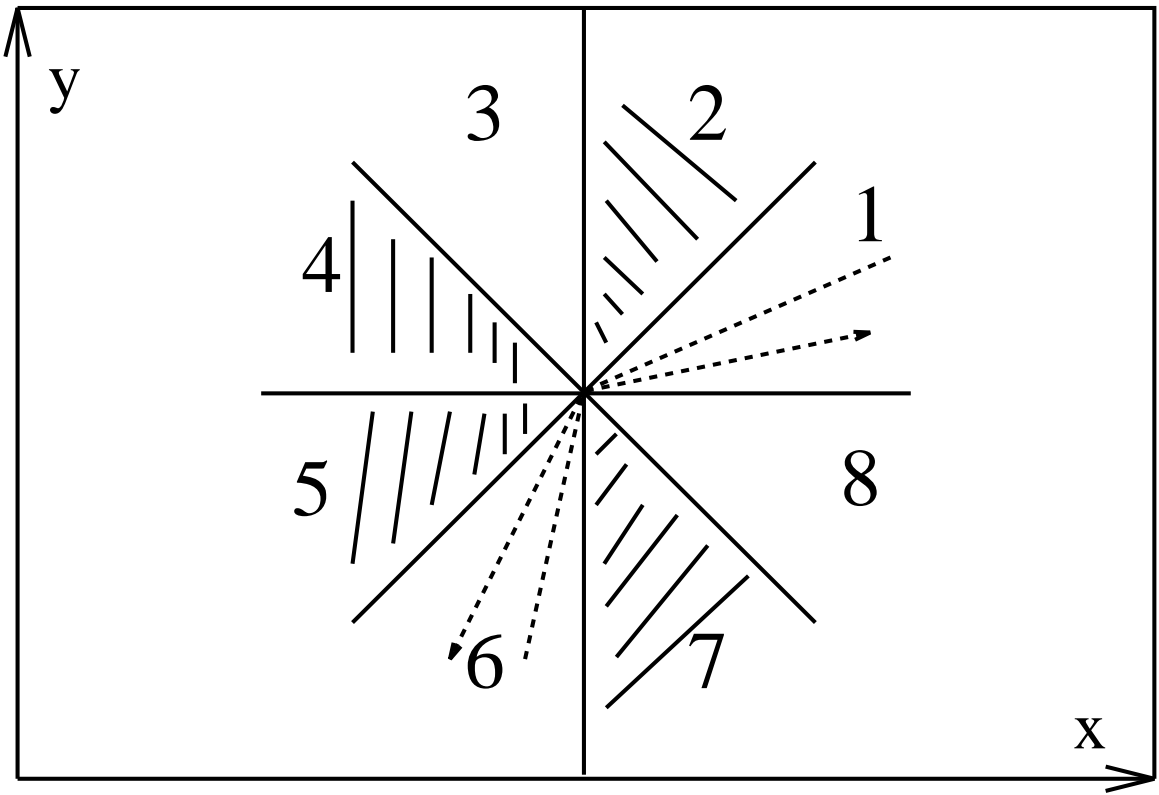

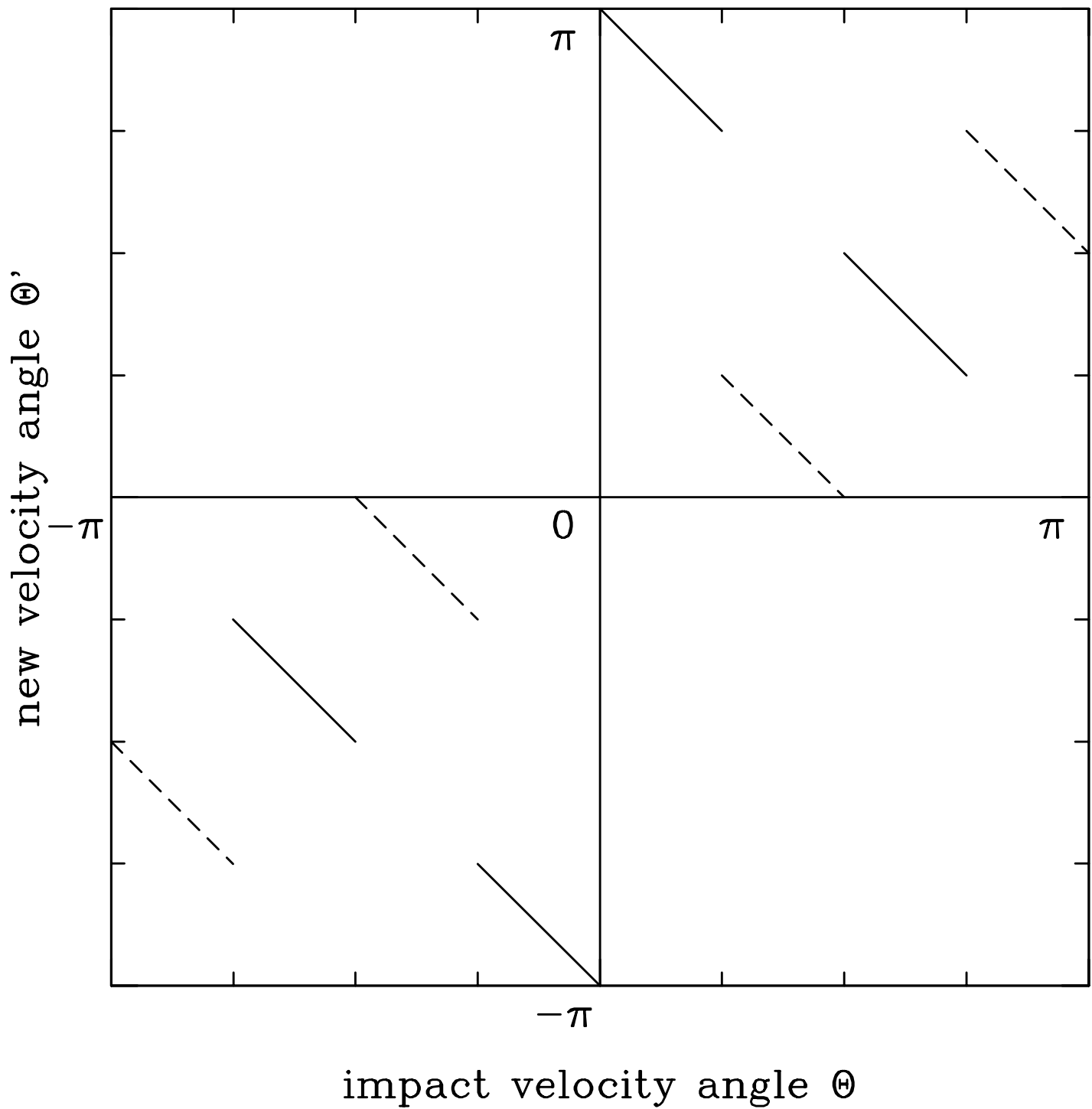

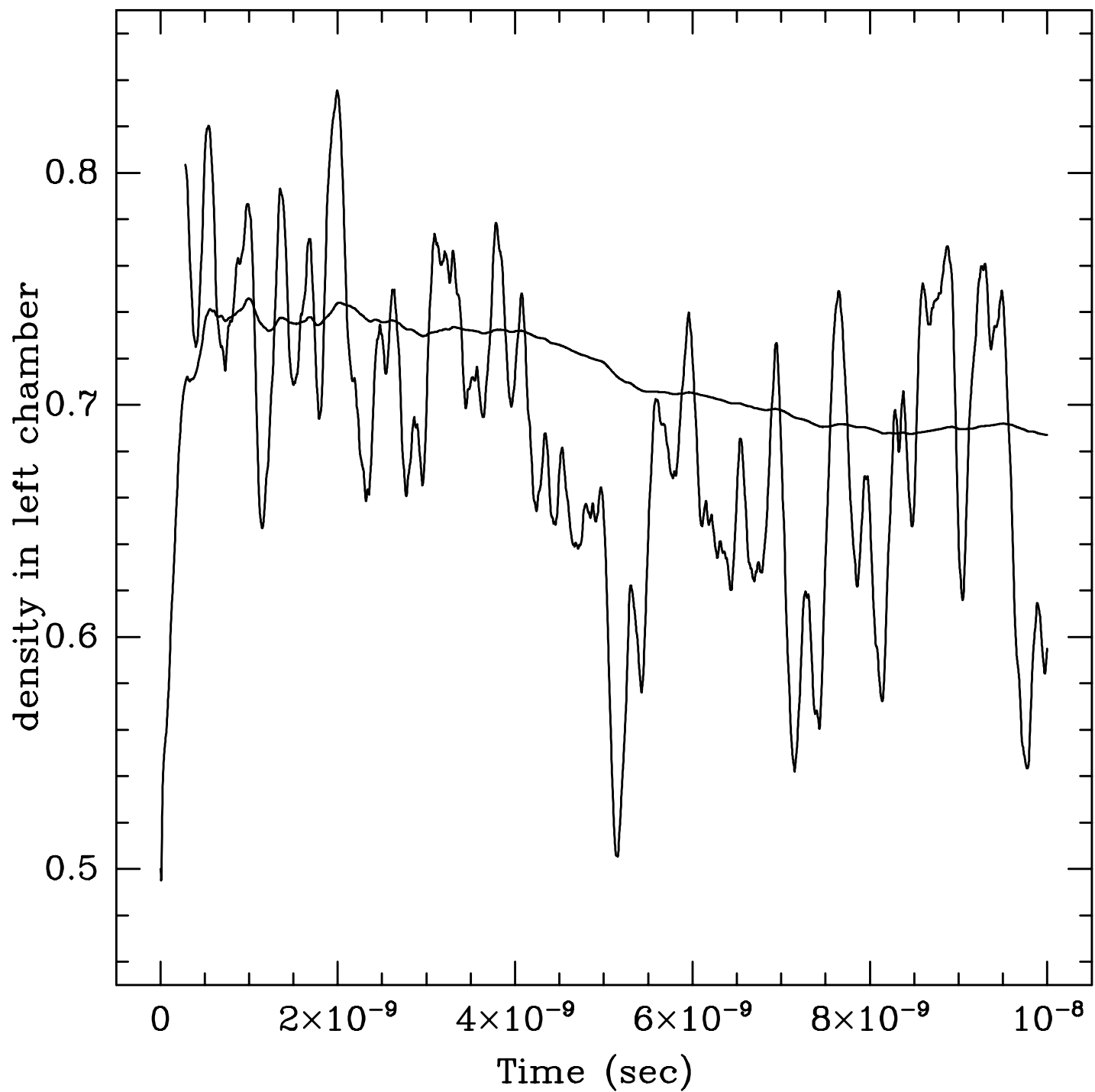

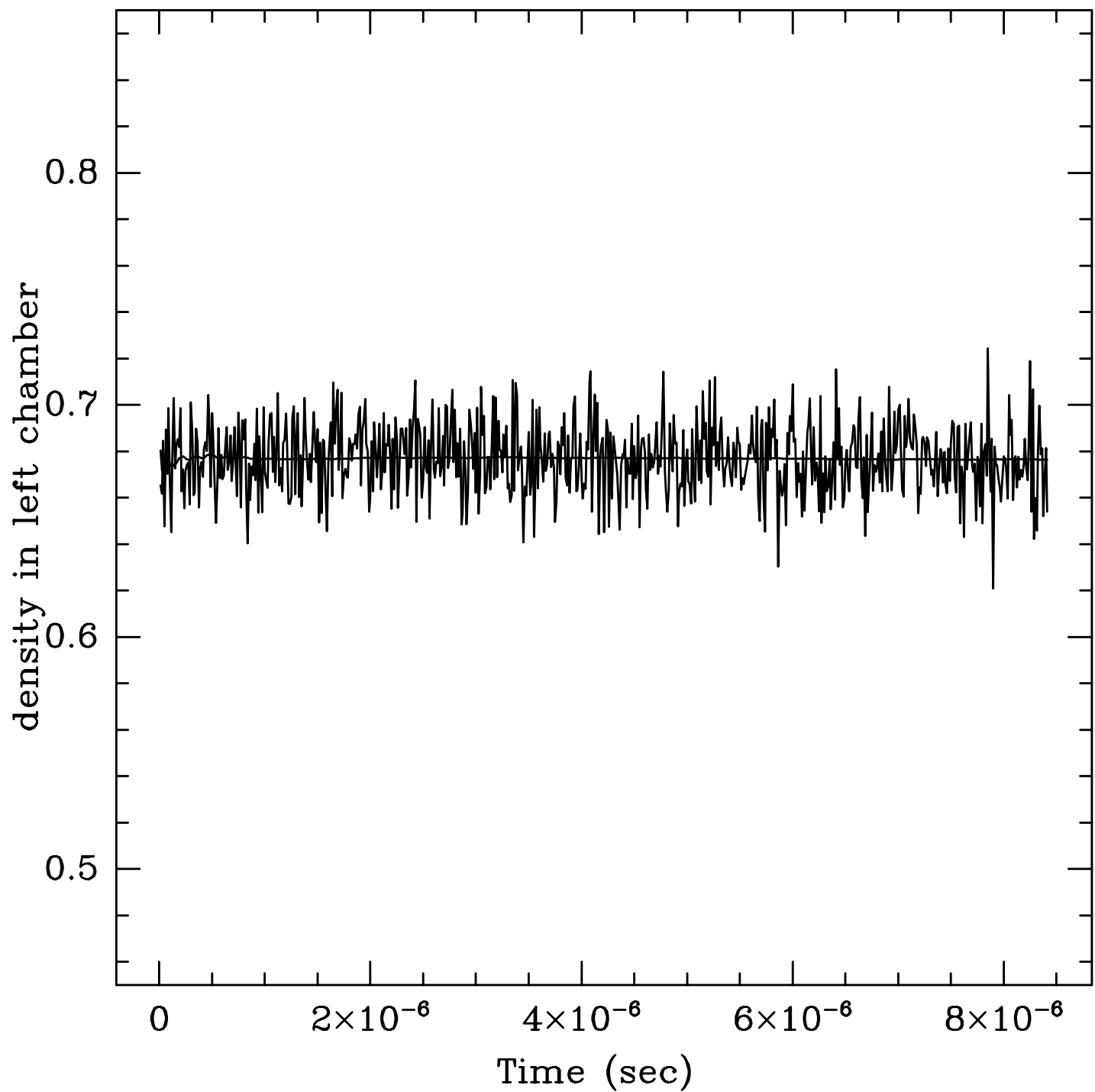

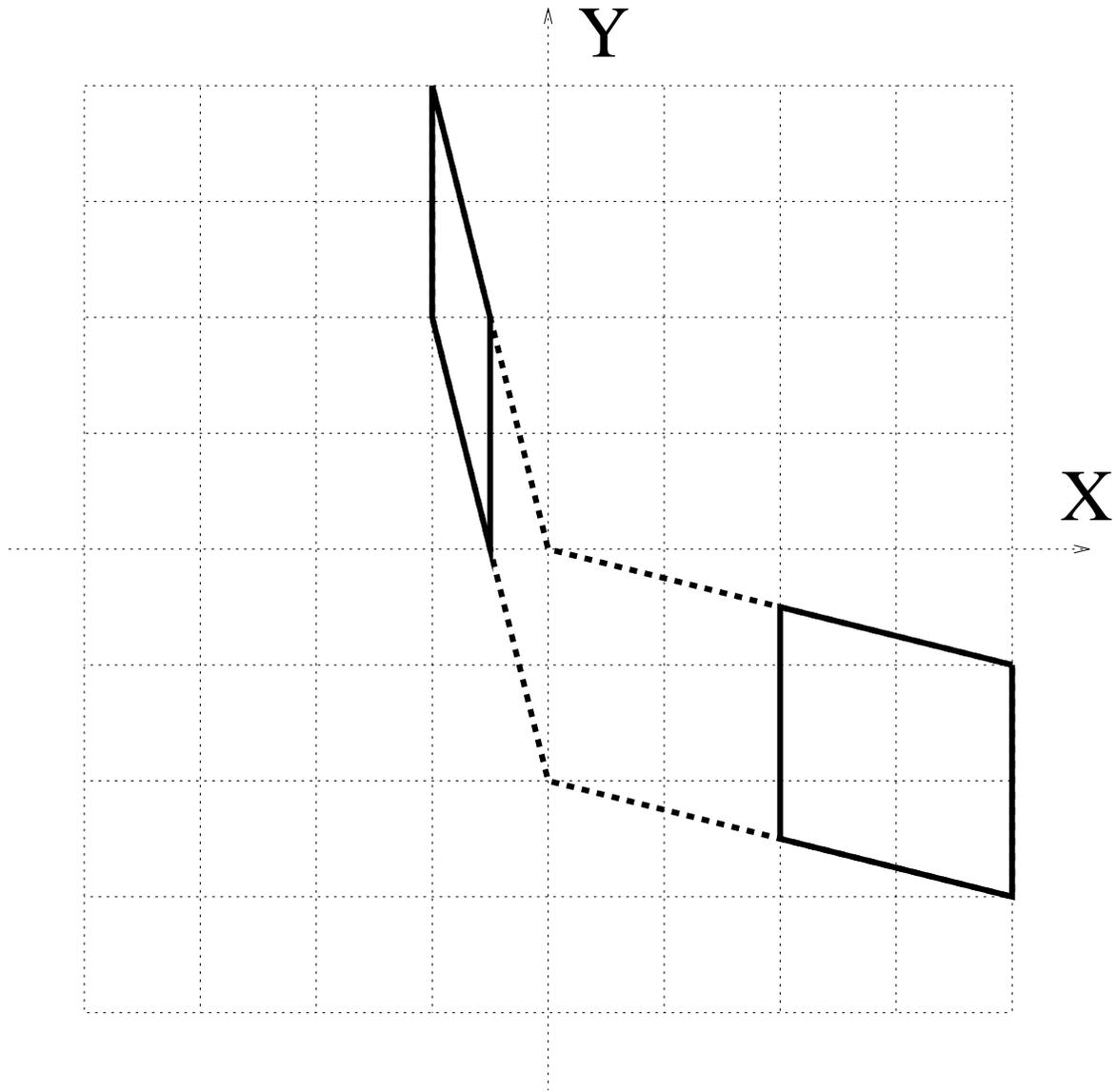

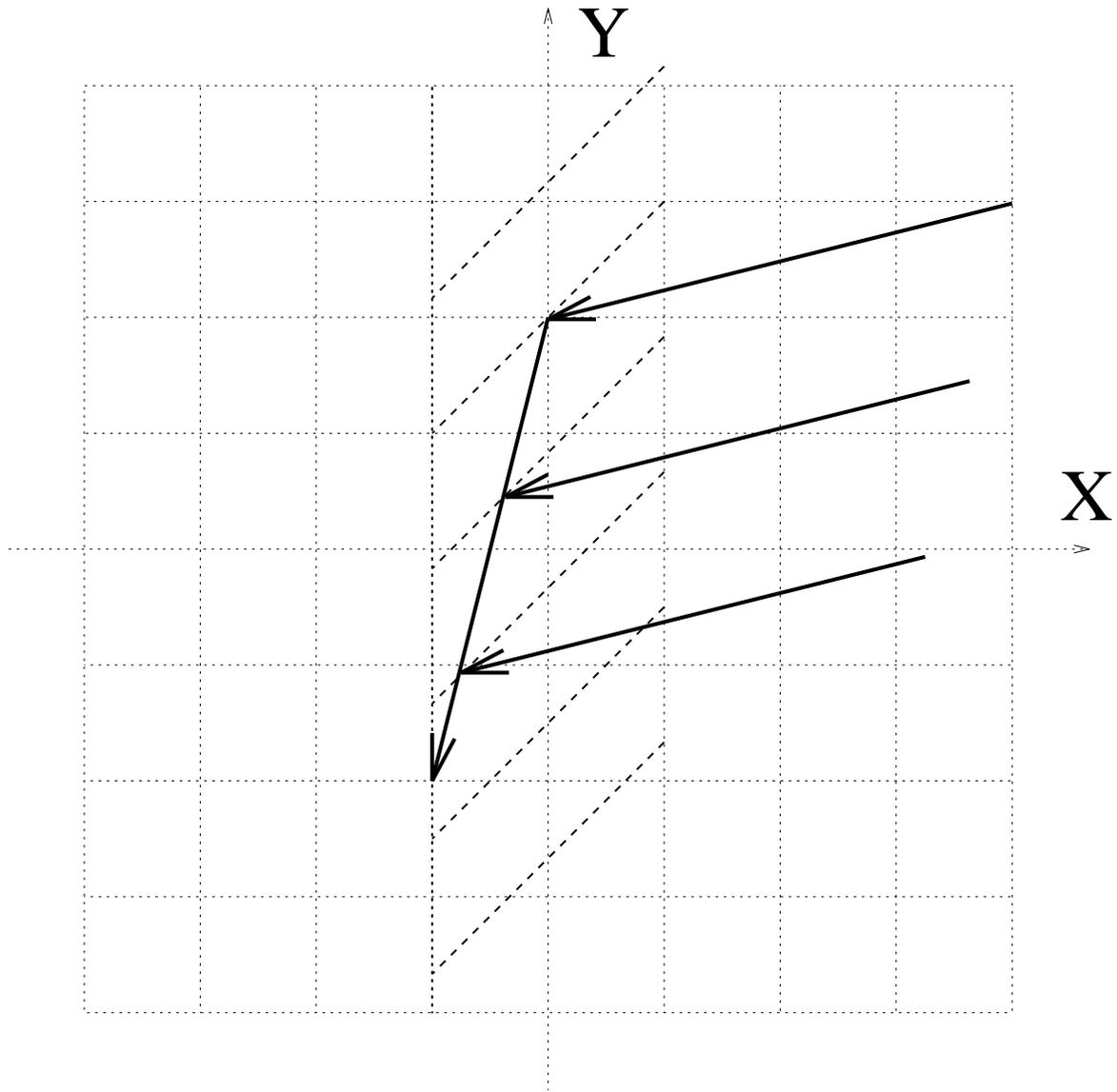

# The cost of compressible dynamics, time-reversibility, Maxwell's demon, and the second law

P. A. Skordos

*Center for Nonlinear Studies*
*Los Alamos National Lab, B258, Los Alamos, NM 87545*
*and*
*Massachusetts Institute of Technology*
*545 Technology Square, NE43-432, Cambridge, MA 02139*

May 14, 1993

**Abstract**

A tantalizing version of Maxwell's demon is presented which appears to operate reversibly. A container of hard core disks is separated into two chambers of equal volume by a membrane that selects which disk can penetrate depending on the disk's angle of incidence. It is shown that the second law of thermodynamics requires the incompressibility of microscopic dynamics or an appropriate energy cost for compressible microscopic dynamics.

# 1  Introduction

The second law of thermodynamics is usually attributed to the fact that states of maximum disorder in a statistical system have the largest probability of occurrence among all possible states. The microscopic dynamics of a statistical system are assumed to be of minor importance. However the microscopic dynamics can not be arbitrary, and they must satisfy certain conditions for the second law to hold. In particular they must conserve phase space volume. Hamiltonian dynamics conserve phase space volume according to Liouville's theorem. However the condition of incompressibility is more general than the condition of Hamiltonian dynamics, and it is important to discuss incompressibility and the second law directly without the aid of Hamiltonians.

This paper examines the relation between compressible microscopic dynamics and the second law of thermodynamics using a *time-reversible system of hard core disks and a membrane*. The possibility that a Maxwell's demon could imitate the membrane of our system in a reversible manner is also explored, and it is shown that a Maxwell's demon can only approximate the membrane using irreversible operations. The membrane selects which disk can penetrate depending on the disk's angle of incidence, and creates a density difference between two chambers initialized to have equal density. The analysis of our system confirms that the second law of thermodynam-ics requires the incompressibility of microscopic dynamics or an appropriate energy cost for compressible microscopic dynamics [1].

The next section describes our system and explains how the membrane creates a density difference between the two chambers. Section 3 estimates the density difference theoretically, and section 4 confirms the theoretical estimate using a computer simulation. Section 5 explains how the membrane challenges the second law of thermodynamics, and discusses the evolution of the system of disks in phase space. Section 6 shows that the membrane compresses the phase space of the disks, and section 7 demonstrates that if the membrane interaction obeyed incompressible dynamics, then it would not create a density difference (for piecewise differentiable maps).

Section 8 reviews another system that obeys compressible dynamics: a pump of disks consisting of a microscopic trapdoor and a cooling mechanism. Section 9 analyzes the membrane as a Maxwell's demon that interacts with the disks based on information about the disks. The new demon uses a tennis racket to bounce the disks and to select which disk can penetrate the membrane depending on the disk's angle of incidence.

## 2   Description of the Model

Our system is shown in figure 1. It consists of a box containing hard core disks, and a membrane that separates the box into two chambers of equal volume. The hard core disks move and collide with each other elastically. The membrane interacts with each incident disk according to the following equation, where $V_1, V_2$ are the velocity components of the disk before the interaction and $V_1', V_2'$ are the velocity components of the disk after the interaction.

$$
\begin{aligned}
+45: & \left\{ \begin{array}{rcl} V_1' &=& V_2 \\ V_2' &=& V_1 \end{array} \right\} \quad \text{if} \quad \left\{ \begin{array}{l} (V_1 < 0 \;\; V_2 < 0 \;\; |V_1| > |V_2|) \text{ or} \\ (V_1 > 0 \;\; V_2 > 0 \;\; |V_1| < |V_2|) \end{array} \right\} \\
-45: & \left\{ \begin{array}{rcl} V_1' &=& -V_2 \\ V_2' &=& -V_1 \end{array} \right\} \quad \text{if} \quad \left\{ \begin{array}{l} (V_1 < 0 \;\; V_2 > 0 \;\; |V_1| > |V_2|) \text{ or} \\ (V_1 > 0 \;\; V_2 < 0 \;\; |V_1| < |V_2|) \end{array} \right\} \quad (1) \\
+90: & \left\{ \begin{array}{rcl} V_1' &=& -V_1 \\ V_2' &=& V_2 \end{array} \right\} \quad \text{otherwise}
\end{aligned}
$$

The labels $+45$, $-45$, and $+90$ are motivated by the discussion of section 9 that views the membrane as a Maxwell's demon playing tennis with the disks. The above equation says that an incident disk in octants 2,4,5, and 7 (refer to figure 2) reverses the x-component of its velocity when it hits the membrane and is not allowed to penetrate. It also says that an incident disk in the remaining octants penetrates the membrane by being deflected toward the x-axis when coming from the left, and away from the x-axis when coming from the right. The speed and the kinetic energy of the disk are conserved.

Equation 1 is illustrated in Figure 2. The vertical solid line in figure 2 denotes the membrane, and the other solid lines denote a division of the plane into octants. The two dashed lines inside the first and sixth octants (counting counterclockwise) denote trajectories that are deflected and penetrate the membrane according to the case +45 of equation 1. The transformation of the velocity takes place instantaneously when the center of an incident disk reaches the membrane.

Another way of examining equation 1 is to rewrite the equation as a map of the velocity angle (impact angle). This is shown in figure 3 where $\Theta$ is the impact velocity angle and $\Theta'$ is the transformed velocity angle after the membrane interaction has occurred. Both $\Theta$ and $\Theta'$ range from $-\pi$ to $\pi$. The map consists of eight line segments which can be interpreted as follows: A completely transparent membrane is a straight line of slope "1" passing through the origin, and a completely impenetrable membrane (mirror-reflecting) is a line of slope "$-1$" moved upwards $\pi$ radians away from the origin and wrapped around periodically. The eight line segments of figure 3 can be interpreted as breaking a line of slope "$-1$" (mirror-reflecting) and shifting some of the broken pieces up and down. The broken pieces that are shown dashed in figure 3 correspond to impact angles that penetrate the membrane. The broken pieces that are shown solid in figure 3 correspond to impact angles that are mirror-reflected back.

The membrane of equation 1 leads to a large density difference between the two chambers. The success of the membrane depends on the thermal motion of the disks and the impact rate of disks hitting the membrane.

Assuming that the velocities of the disks are distributed isotropically inside the container, it follows from geometrical considerations that the impact rate of disks hitting the membrane is a cosine of the impact angle in absolute value,

$$\text{impact-rate} \quad \propto \quad |\cos \Theta| . \qquad (2)$$

The membrane exploits this cosine distribution of impact angles by allowing disks with "high rate" impact angle to penetrate from the right, and allowing disks with "low rate" impact angle to penetrate from the left. To achieve reversibility, the remaining impact angles ("low rate" from the right and "high rate" from the left) are blocked and do not penetrate the membrane. They are simply reflected back. Furthermore, each "high rate" angle from the right is rotated onto a "low rate" angle when the disk penetrates the membrane, and vice versa.

The membrane interaction of equation 1 makes the membrane more permeable from the right side than from the left. This leads to an excess flux of disks from the right side, and creates a density difference between the two chambers. When the density difference reaches an equilibrium value (which is calculated in the next section), the fluxes of disks between the two sides of the membrane become equal.

## 3  Estimate of the Density Difference

At equilibrium the fluxes of disks from the left and from the right side of the membrane are equal to each other. In other words if $N_L$ is the normalized density in the left chamber, and $P_{(L \to R)}$ is the probability of an *individual disk* to penetrate the membrane coming from the left, we require that

$$N_L \ P_{(L \to R)} \ = \ N_R \ P_{(R \to L)} \ . \tag{3}$$

We can estimate $P_{(L \to R)}$ using the fact that among all the disks that strike the membrane from the left only those with trajectories in the third and sixth octants of figure 2 are allowed to penetrate. In particular, these disks have impact angles in the intervals $(\pi/4, \pi/2)$ and $(\pi, 5\pi/4)$. In addition, we assume that the probability of an individual disk to strike the membrane varies as the cosine of the impact angle and is independent of the density in each chamber. Although the total impact rate depends linearly on the density of the disks, the probability of an individual disk to reach the membrane is independent of the density to a first approximation. Thus we get,

$$P_{(L \to R)} \ \simeq \ 2 \ C \ \int_{\pi/4}^{\pi/2} \cos \theta \ d\theta \ \simeq \ 2 \ C \ 0.3 \ , \tag{4}$$

for some normalization constant $C$; and similarly,

$$P_{(R \to L)} \ \simeq \ 2 \ C \ \int_{4\pi/4}^{5\pi/4} \cos \theta \ d\theta \ \simeq \ 2 \ C \ 0.7 \ , \tag{5}$$

which gives

$$N_L \ \simeq \ 0.7 \ \ \text{and} \ \ N_R \ \simeq \ 0.3 \ . \tag{6}$$

In other words the system of membrane and disks reaches equilibrium when the fluxes of disks from the left and from the right side of the membrane are balanced, and this happens when the normalized density is approximately 0.7 in the left chamber and approximately 0.3 in the right chamber.

## 4 Simulation Results

To check the theoretical results of the previous section, a two-dimensional system of hard core disks with a membrane has been simulated. The computer program used in these simulations is the same program as the one described in detail in reference [2] with a few modifications to simulate the membrane. In particular, when the center of an incident disk reaches the membrane, the disk's velocity is transformed according to equation 1.

In the experiments reported below forty disks are used. The size of each chamber is $24.3 \times 10^{-13} cm^2$ (equal size chambers), and the disk radius is $3 \times 10^{-8} cm$. These numbers give a mean free path of the order of $10^{-6} cm$ which is approximately the length of each chamber. The average speed of each disk is $3.56 \times 10^4 cm/sec$.

In figure 4 we can see that the time average of the number of disks in the left chamber increases from an initial value of 0.5 (normalized) to a value of 0.7 as a result of the membrane interaction. The smooth curve of figure 4 plots the cumulative time average of the number of disks in the left chamber, which approaches a steady value as the averaging time increases. The noisy curve of figure 4 plots a sequence of running averages of the number of disks in the left chamber (each one taken over $1.25 \times 10^{-10} sec$), and provides an indication of the density fluctuations for the chosen system parameters.

Figure 5 shows the same quantities as figure 4, but examines them on a much longer time scale. The running averages of figure 5 are based on longer time intervals ($25 \times 10^{-10} sec$), so the size of fluctuations is accordingly reduced. The cumulative time average of the number of disks in the left chamber is a straight line that intersects the y-axis at the value of 0.68 (normalized). This corresponds to 0.69 density (normalized) if we take into account that the unequal number of disks in each chamber changes the available area. The simulation results are in good agreement with the theoretical estimate of 0.7 normalized density difference of section 3.

## 5  The Loss of Ergodicity

Our membrane appears to create a density difference without doing any work. In particular, it appears to establish the equivalent of a lower potential energy in the left chamber that creates a density difference in order to equalize the potential difference between the two sides. This can not be correct however. If we cut a hole in the membrane, a *return flux* of disks will result through the hole which can be used to convert thermal energy into useful work in violation of the second law of thermodynamics.

In order to rescue the second law of thermodynamics we can follow a number of different approaches. The first approach is to assume that the membrane *can not be cut*, that it must be a closed surface. In nature for example a chemical potential difference exists at the interface between two different materials such p-n semiconductors (see [3]). There is nothing in our model of the membrane however that implies that the membrane is an interface between two different materials and that it can not be cut. So we have to look elsewhere for an explanation of why the membrane can not violate the second law of thermodynamics. We start by examining the evolution of the disks in phase space.

A system of $N$ disks in two dimensions can be represented as a $4N$ vector $(\ldots, X_i, Y_i, U_i, V_i, \ldots)$ of real numbers in $R^{4N}$ phase space. This $4N$ vector is the *representative point* of the system, and it specifies exactly the positions and velocities of the particles in the system at any given time (all particles have equal mass). As the system evolves in time, the representative point moves inside a constant energy subsurface of the $R^{4N}$ phase space because the total energy is conserved. The total linear momentum is not conserved because wall collisions reflect a disk's momentum $U'_i = -U_i$ or $V'_i = -V_i$ and because membrane interactions permute and/or reflect a disk's momentum according to equation 1. The total linear momentum is only conserved in a time average sense, and this has been checked by computer simulations.

The membrane and disks system is not ergodic because it does not spend equal times in equal regions of the accessible phase space $\Omega$ (for example see [4, p.68]). If the representative point of the system visited regions of $\Omega$ that have more disks in the left chamber as often as regions of $\Omega$ that have more disks in the right chamber, then the time average density would be equal in the two chambers. Instead, the membrane and disks system approaches irreversibly a state of 0.7/0.3 density difference independent of initial conditions.

In general there are many ways in which a system can fail to be ergodic. A trivial way to lose ergodicity in the context of billiard balls is to remove all interactions between the disks. Then the disks can not see each other and bounce between the walls of the container undisturbed, and the system does not attain a Maxwellian velocity distribution. This is a trivial example that shows that collisions between disks are necessary for modeling ideal gas. Binary collisions give rise to chaotic dynamics and allow a system of hard disks to explore fully its phase space.

Binary collisions are not enough to guarantee ergodicity however. A system based on binary collisions and some other dynamics can fail to be ergodic if the additional dynamics introduce an attractor in phase space. In particular this can occur when the evolution map $M : \Omega \to \Omega$ does not conserve measure in phase space; that is the Jacobian determinant of the evolution map is not equal to one (absolute value). The simplest example of a map that does not conserve measure is a two-to-one map, which means that two distinct representative points in $\Omega$ are mapped onto the same point. In physical space it means that two distinct trajectories of disks are mapped onto the same trajectory. An example is the one-way valve Maxwell's demon described in section 9.

It turns out that the membrane and disks system also compresses phase space volume. The following three sections analyze this compression and demonstrate that the second of thermodynamics requires the incompressibility of microscopic dynamics or an appropriate energy cost for compressible microscopic dynamics.

# 6  Compression of Phase Space Volume

To examine the compressibility of dynamics, we consider a membrane system that contains only one disk for simplicity. In other words, we have the same container and membrane as shown in figure 2 and we have a single disk inside. The phase space $\Omega$ is three dimensional $X, Y, \Theta$ where $X, Y$ is the position of the disk and $\Theta$ is the velocity angle. We examine the compressibility of dynamics using this one disk system.

It is easy to see that during free motion and collisions with the wall, the evolution of the system conserves phase space volume. The question is what happens during the membrane-disk interaction. Referring to figure 3 we see that the transformation of the velocity angle of an incident disk is a linear map that has slope minus one, which suggests that phase space volume is conserved. This is incorrect however. In order to calculate the compressibility of dynamics we must consider the evolution of all three $X, Y, \Theta$ dimensions of the phase space together, and not only the $\Theta$ dimension that is shown in figure 3.

For concreteness we look at the time evolution of an infinitesimal phase space volume $\omega(X_1, Y_1, \Theta_1)$ centered at the representative point $X_1, Y_1, \Theta_1$. We assume that the membrane is located at $X = 0$, and that $X_1$ is near the membrane with $X_1 > 0$, and that $\Theta_1$ is inside the interval $3\pi/4 < \Theta_1 < \pi$. We assume that after a time interval of 1.0 (in appropriate units) every point $X, Y, \Theta$ in the volume $\omega(X_1, Y_1, \Theta_1)$ has penetrated the membrane from the right and moved to the left of the membrane. We denote by $X', Y', \Theta'$ the image of $X, Y, \Theta$ under the evolution map, and we have the following equation,

$$X' = X + \cos\Theta \, \Delta t_c - \sin\Theta \, (1 - \Delta t_c) \qquad (7)$$

where $\Delta t_c$ is the time it takes for point $X, Y, \Theta$ to move to the membrane and is equal to $-X/\cos\Theta$. Therefore,

$$\begin{aligned} X' &= -\sin\Theta - X\sin\Theta/\cos\Theta \\ Y' &= Y - X(1+\sin\Theta/\cos\Theta) - \cos\Theta \\ \Theta' &= 3\pi/2 - \Theta \end{aligned} \quad (8)$$

To check whether the evolution map compresses the volume $\omega(X_1, Y_1, \Theta_1)$ we evaluate the Jacobian determinant of the above equations. We find,

$$\begin{vmatrix} -\sin\Theta/\cos\Theta & 0 & -\cos\Theta - X/\cos^2\Theta \\ -1-\sin\Theta/\cos\Theta & 1 & \sin\Theta - X/\cos^2\Theta \\ 0 & 0 & -1 \end{vmatrix} = \sin\Theta/\cos\Theta = \tan\Theta$$

(9)

For points $X, Y, \Theta$ inside the volume $\omega(X_1, Y_1, \Theta_1)$, the angle $\Theta$ is inside the interval $3\pi/4 < \Theta < \pi$. Thus, the Jacobian determinant is always less than one. In other words the evolution map compresses phase space volume when a disk penetrates the membrane from right to left.

Figure 6 shows geometrically how the compression of phase space volume occurs. The figure is confined to two dimensions for graphical reasons. The two rectangles shown in solid lines correspond to phase space volumes that are mapped onto each other under the evolution map — they correspond to $\omega(X, Y, \Theta)$ with the angle $\Theta$ chosen constant for all points. The vertical line $X = 0$ of figure 6 corresponds to the membrane. It is easy to see that the edges of the two rectangles (the original rectangle and its image under the evolution map) are equal between the two rectangles, but the corresponding angles are not equal, and hence the areas of the two rectangles are not equal. Phase space volume is compressed when penetrating the membrane from right to left and expanded when penetrating from left to right.

# 7 Other Membrane Maps

The significance of compressible dynamics can be appreciated if we attempt to find a new membrane map that would result in a density difference while preserving phase space volume. It turns out that this is not possible, at least for piecewise differentiable maps. To see this, we seek a map $f(\Theta)$ mapping the impact velocity angle $\Theta$ to a new velocity angle $f(\Theta)$ so that the condition of incompressibility (Jacobian determinant unity) is satisfied. Repeating the above steps, equations 8 and 9 using $f(\Theta)$, we find

$$f(\Theta) = \sin^{-1}(\pm \sin \Theta + C) \qquad (10)$$

where $C$ is an arbitrary constant. If $C$ is zero, $f(\Theta)$ corresponds to a transparent membrane (i.e. no membrane at all) or a completely impenetrable membrane (i.e. a wall). If $C$ is non-zero, $f(\Theta)$ corresponds to a membrane that maps velocity angles in a non-linear fashion. In analogy with the membrane of equation 1, we apply the non-linear map $f(\Theta)$ of equation 10 to a selected region of velocity angles and block the remaining angles. In this way we hope that the probabilities of penetrating the membrane from the left and from the right will be unequal (see section 2). A simple calculation however shows that the probabilities of penetrating the membrane from the left and from the right will be equal for all possible choices of the constant $C$ in equation 10.

For example let us suppose that $(\pi, \pi - d)$ is an interval of velocity angles that penetrate the membrane from the right side, and $(f(\pi - d), f(\pi))$ is the image of $(\pi, \pi - d)$ under the membrane map $f(\Theta) = \pi - \sin^{-1}(\sin \Theta + C)$ where $C$ is a positive constant $C <= 1 - \sin d$. Also let us assume that the remaining velocity angles are blocked and do not penetrate the membrane. Then the probability of an individual molecule to penetrate the membrane from the right (see section 3) is,

$$P_{(R \to L)} = \int_{\pi-d}^{\pi} |\cos \theta| d\theta , \tag{11}$$

and the probability to penetrate from the left is,

$$P_{(L \to R)} = \int_{f(\pi-d)}^{f(\pi)} |\cos \theta| d\theta . \tag{12}$$

An elementary integration gives

$$\begin{aligned} P_{(R \to L)} &= \sin(\pi - d) = \sin d , \\ P_{(L \to R)} &= \sin f(\pi - d) - \sin f(\pi) = \\ &\sin[\sin^{-1}[\sin(\pi - d) + C]] - \sin[\sin^{-1}[\sin \pi + C]] = \\ &\sin d . \end{aligned} \tag{13}$$

Hence the probabilities of penetrating the membrane from the right and from the left are equal to each other. Therefore a membrane map (piecewise differentiable) that obeys incompressible dynamics can not create a density difference.

The above discussion confirms that the membrane of equation 1 achieves a density difference by compressing the phase space of the disks. The density difference achieved by the membrane can be brought in accordance with the second law of thermodynamics by assuming that the membrane does a certain amount of work in order to compress the phase space of the disks. In other words, the second law of thermodynamics requires the incompressibility of microscopic dynamics or an appropriate energy cost for compressible microscopic dynamics.

## 8  Microscopic Rectifiers

Another system that obeys compressible microscopic dynamics with an energy cost is a pump of disks consisting of a microscopic trapdoor and external cooling. Such a pump is often used to explain thermalization [2, 8, 9]. In this section we examine the pump in terms of phase space compression.

A microscopic trapdoor *without external cooling* is an example of a system that is designed to extract work from the thermal motion of disks by rectifying spontaneous variations in density in a system of disks. Such a trapdoor is also called a *microscopic rectifier*, and it does not succeed as explained in references [2, 8, 9] because the rectifying mechanism becomes thermalized and starts moving randomly in every possible way. In order to succeed the rectifying mechanism must be kept at a lower temperature than the system of disks.

Interestingly the prevention of thermalization by a cooling process compresses phase space volume. To see how compression of phase space volume occurs when a microscopic rectifier is cooled, we examine the trapdoor system of reference [2]. For simplicity we consider only one disk at first. We denote by $x, y, u, v$ the coordinates and velocity of the disk, and by $X, U$ the position and velocity of the trapdoor. We assume that the cooling operation of reference [2] occurs at time $T_c$ for $0 < T_c < 1$ (in appropriate units of time), and we consider the evolution of the system between times 0 and 1. The new state is given by the following equations,

$$\begin{aligned}
x' &= x + u\,T_c + \eta\,u\,(1 - T_c) \\
y' &= y + v\,T_c + \eta\,v\,(1 - T_c) \\
u' &= \eta\,u \\
v' &= \eta\,v \\
X' &= X + U\,T_c + \epsilon\,U\,(1 - T_c) \\
U' &= \epsilon\,U
\end{aligned} \qquad (14)$$

where $0 < \epsilon < 1$ is the cooling parameter (a fixed number), and $\eta$ is chosen so that the total energy of the system is conserved. If $m, M$ are the masses of the disk and the trapdoor, the parameter $\eta$ is given by the formula,

$$\eta = \sqrt{1 + \frac{MU^2}{m(u^2+v^2)}(1-\epsilon^2)} \quad . \tag{15}$$

After some algebra we find that the Jacobian determinant of the evolution equation 14 is equal to $\epsilon$. In other words, phase space volume is compressed by a factor of $\epsilon$ during a cooling operation. A similar calculation for a system containing a large number of disks, for example $N$ disks, gives the following Jacobian determinant,

$$\text{Jacobian} = \epsilon \left(1 + (1-\epsilon^2)\frac{MU^2}{\sum_{i=1}^{N} m(u_i^2+v_i^2)}\right)^{(N-1)}, \tag{16}$$

which reduces to $\epsilon\left(1 + \gamma(1-\epsilon^2)\right)$ if we assume that the total energy of the disks is much larger than the energy of the trapdoor, and that $\gamma$ is the ratio of the energy of the trapdoor to the average energy of the disks. Further if the trapdoor is colder than the disks, the energy ratio $\gamma$ is a small number, and the Jacobian determinant is approximately equal to $\epsilon$. Hence, cooling the trapdoor is accompanied by compression of phase space volume.

A cooled rectifier is simply a pump and is not a threat to second law of thermodynamics [2]. The external cooling does work on the system and is responsible for compressing the phase space of the trapdoor and disks. When the cooled rectifier is enclosed in a larger system that is closed and isolated (for example if hot and cold reservoirs of finite heat capacity are used to perform the cooling), then the extended phase space evolves incompressibly and heat can only be converted to work while the system is approaching equilibrium. The membrane and disks system can also be viewed as a pump if we assume that the membrane does work in order to compress the phase space of the disks.

## 9 Maxwell's demon

An alternative way of bringing the membrane in accordance with the second law of thermodynamics is to view the membrane as a Maxwell's demon that interacts with the disks according to information about each incoming disk. For this purpose we review first the traditional one-way valve Maxwell's demon.

Maxwell's demon is an imaginary being (or device) that operates a microscopic door between two chambers containing disks [5, 2, 6, 7]. The simplest version of the demon opens the door when a disk is coming from the right, and closes the door when a disk is coming from the left. The demon's operations lead to a density difference between two chambers, and the density difference can be used to extract work from the thermal motion of the disks. If the demon could operate in a complete cycle dissipating less energy than the work that can be extracted after the demon has finished its operations, then the second law of thermodynamics would be violated. This conundrum has inspired a large volume of literature aimed at exorcising Maxwell's de-mon [6].

A popular way of exorcising Maxwell's demon (see [10]) is to assume that the demon operates its trapdoor according to information about each incoming disk, and to observe that the demon must erase the information that it has about the present disk in order to process the next disk. The erasure of information is irreversible because the demon maps two distinct disk trajectories onto the same trajectory every time it interacts with a disk by opening and closing its trapdoor. In particular after an interaction it is impossible to distinguish whether the disk came from the opposite chamber or whether the disk bounced off the demon's door.

It is postulated that the irreversible erasure of information must be accompanied by a minimum amount of entropy production which is $ln(2)$ in the case of the one-way valve Maxwell's demon [10]. Thus the demon is brought into accordance with the second law because the reduction of entropy achieved by the demon is counterbalanced by an equal amount of entropy production that is necessary to implement the demon's irreversible operations.

The membrane of equation 1 can also be viewed as a Maxwell's demon by imagining that the membrane-disk interactions are the result of a *demon playing tennis with the disks*. The demon moves a tiny racket up or down so as to intercept each incoming disk at the membrane line. In addition, the demon orients the racket in one of three possible orientations $+45$, $-45$, and

+90 degrees, so as to reflect the incoming disk according to equation 1 (also see figure 2).

The tennis demon differs from the one-way valve Maxwell's demon in that the racket of the tennis demon must be positioned along a continuum of locations depending on where the disk intersects the membrane line. In contrast, the door of the one-way valve Maxwell's demon must be positioned in one of two locations, closed or open. Another difference between the tennis demon and the one-way valve Maxwell's demon is that the tennis demon discards no information about the disks: The disk trajectories evolve bijectively so that the state of a disk (position, velocity) after a racket collision determines completely the state of the disk before the racket collision. In contrast, the one-way valve Maxwell's demon maps two disk trajectories onto the same trajectory every time it interacts with a disk.

How can we then exorcise the tennis demon using information ideas? In section 6 we saw that the evolution of the disks compresses phase space volume. Now we may notice that the compression of phase space volume occurs because the demon chooses the location of the racket as a function of the incoming disk's position. In contrast, a racket that is positioned at a *fixed location* does not cause any compression of phase space volume. Therefore the problem must lie in choosing the racket locations.

Suppose we consider a tennis demon that can only position the racket at a discrete number of fixed locations along the membrane line. In particular we assume that the possible racket locations are uniformly spaced, and that the distance between successive locations is equal to $\Delta x$. If a disk trajectory does not fall exactly on one of the racket locations, the demon picks the nearest location possible. With the help of a diagram (see figure 7) we can see that no matter how small (but non-zero) the spacing $\Delta x$ is, there are always disk trajectories that are mapped on top of each other. The discrete tennis demon does not compress the phase space of the disks in the continuous manner of section 6, but it does compress the phase space of the disks because of the many-to-one map of disk trajectories.

The discrete tennis demon operates with finite information about each incoming disk, but it erases information irreversibly. The irreversible many-to-one map of disk trajectories disappears only in the limit of $\Delta x$ going to zero (the spacing between racket locations). In this limit the discrete compression of phase space volume is replaced by the continuous compression of phase space volume of section 6. Moreover the reversibility of trajectories that is achieved in the limit comes at the expense of requiring the demon to operate with infinite information. Because Maxwell's demon can only operate with finite information (we can think of it as a microscopic computer), it follows that the tennis demon can not imitate the membrane of equation 1 reversibly. A tennis demon can only approximate the membrane of equation 1 using irreversible operations.

Figure 1: A system of disks and a membrane can be viewed as a Maxwell's demon (see text). The membrane, displayed as a dashed line, interacts with the disks according to a time-reversible and energy conserving rule that creates a density difference between the two chambers.

Figure 2: The membrane interaction is shown graphically. The vertical solid line denotes the membrane, and the other solid lines denote a division of the plane into octants. The dashed lines correspond to disk trajectories that penetrate the membrane.

Figure 3: The membrane interaction is shown as a map of the velocity angle. The impact velocity angle $\Theta$ is mapped to the new velocity angle $\Theta'$. Both angles range from $-\pi$ to $\pi$. The dashed line segments correspond to trajectories that penetrate the membrane, while the solid line segments correspond to trajectories that are mirror-reflected back.

Figure 4: The membrane creates a density difference between two chambers initialized to have equal density. The smooth curve plots the cumulative time average of the number of disks in the left chamber, which increases from an initial value of 0.5 to approximately 0.7. The noisy curve plots a sequence of running averages of the number of disks in the left chamber, each one taken over $1.25 \times 10^{-10} sec$.

Figure 5: The same quantities as in figure 4 are examined on a much longer time scale. The running averages are based on time intervals of $25 \times 10^{-10} sec$.

Figure 6: The membrane compresses the phase space of the disks. A region of phase space that corresponds to disks with identical velocity angle is compressed when the points penetrate the membrane from right to left.

Figure 7: A finite demon can position its racket at a discrete number of locations that are uniformly spaced along the membrane line. The demon picks the nearest location possible to position its racket, but there are always trajectories that are mapped on top of each other. The y-axis corresponds to the membrane and the solid lines are disk trajectories.